\documentstyle[preprint,aps]{revtex}

\begin{document}
\draft
\title{
Ginzburg-Landau Theory of Josephson Field Effect Transistors}
\author{J.J. Betouras and Robert Joynt}
\address{
Department of Physics and Applied Superconductivity Center\\
University of Wisconsin-Madison\\
Madison, Wisconsin 53706\\}
\author{ Zi-Wen Dong and  T. Venkatesan}
\address{
Center for Superconductivity Research, Department of Physics\\
University of Maryland\\
College Park, Maryland 20742}
\author{Peter Hadley}
\address{
Department of Applied Physics\\
Delft University of Technology\\
P.O. Box  5046, NL-2600 GA Delft\\
The Netherlands\\}
\date{\today}
\maketitle

\begin{abstract}
A theoretical model of high-T$_c$ Josephson Field Effect Transistors 
(JoFETs) based on a Ginzburg-Landau free energy expression whose parameters
are field- and spatially- dependent is developed. This model is used to explain
experimental data on JoFETs made by the hole - overdoped
Ca-SBCO bicrystal junctions (three terminal devices).
The measurements showed a large modulation of the critical current as
a function of the applied voltage due to charge modulation in the bicrystal
junction. The experimental data agree with the solutions of the
theoretical model.  This provides an explanation of the large field effect,
based on the strong suppression of the carrier density near the grain 
boundary junction in the absence of applied field, and the subsequent
modulation of the density by the field.
\end{abstract}
\narrowtext
\newpage

The development of practical high-T$_{c}$ three terminal devices has received
much attention recently. There are many different directions that have
been investigated: the electric field effect transistor,
the flux-flow transistor, quasiparticle injection devices, etc. 
The superconducting field effect transistors (suFETs) with a homogeneous
thin film channel \cite{XiPRL,Man} and JoFETs with a Josephson junction channel 
\cite{Dong1,Dong,Ivanov,Petersen,nakajima,Mannhart} 
are electric field related devices. 
In general, the required electric field for the field effect
in JoFETs is orders of magnitude smaller than the field required in
suFETs with homogeneous thin film.

A basic question that therefore arises is to understand 
the large effect that an electric field has on the transport 
properties in JoFETs. Dong et al. \cite{Dong1,Dong}
showed that there is a 23\% modulation of the critical current on
a grain boundary of a 50 nm thick channel of Sm$_{1-x}$Ca$_{x}$Ba$_{2}$
Cu$_{3}$O$_{y}$ (Ca-SBCO).
Nakajima et al. \cite{nakajima} reported a 5\% modulation of critical current
on a 60 nm thick YBCO grain boundary junction channel which can be 
described by a parallel resistor model.
A modulation of several per cent (maximum 8\% in $I_c$) has been reported by
Petersen et al. \cite{Petersen}
on the transport properties of a less than 32 nm thick YBCO grain boundary
junction channel
both in the normal and superconducting state. 
The most recent JoFET experiments were carried out by Mannhart's group
\cite{Mannhart} where a similar field modulation of $I_c$ 
was observed. However it is still not clear why all suFETs with grain boundary
junction (JoFETs) show much bigger effects than 
suFETs made of homogeneous film under the same applied field. 
Candidate explanations include a weakly coupled SNS model in the dirty limit
\cite{Dong}, a parallel resistor model at high bias current,
and the electromechanical effect in the dielectric layer
at low bias currents \cite{Petersen}. But none of the above
gives a quantitative explanation of the dependence of the critical 
current on the field.
The mechanism responsible
for this large field effect therefore remains an open and 
very important question.

In this paper we attempt to clarify this issue, using a phenomenological
model based on the Ginzburg-Landau (GL) theory of phase transitions. 
A similar model
has been already developed for the case of YBa$_2$Cu$_3$O$_{7-\delta}$
grain boundaries in 
bicrystals \cite{BetourasJoynt}.  The basic result of that study was that
the oxygen depletion can account for a major portion of the change from
weak to strong coupling of grain boundaries,
which is experimentally observed \cite{sue} as the misorientation 
angle is increased.
The modification of the oxygen content leads to the variation of the critical
temperature as a function of distance from the boundary.
The detailed way that this occurs is not well 
understood from a microscopic point 
of view. The phenomenological approach of Ref.\ \cite{BetourasJoynt}
allows one to simulate
the behavior of the system in terms of a few measured parameters and to
calculate electromagnetic properties.
Our aim in this paper is to apply this method to the field efffect
modulations seen in the JoFETs
of Ref.\ \cite{Dong1,Dong}.  
We demonstrate that the JoFET systems can be fit into the same conceptual
framework as the YBa$_2$Cu$_3$O$_{7-\delta}$ grain boundaries.

The experimental measurements \cite{Dong1} were taken on
hole-overdoped Ca-SBCO 
bicrystal junctions with 30\% doping of
Ca ($ x = 0.3$) at 20 K and 4.2 K.  
The junction itself is a grain boundary
with a misorientation of 24$^{\circ}$, created by growing the high-T$_c$
film on an SrTiO$_3$ substrate with such a boundary. A schematic
view of the experimental setup is shown in Fig.\ 1.  
Junctions of this kind in the 
YBa$_2$Cu$_3$O$_{7-\delta}$ system have been shown 
to be oxygen-deficient \cite{sue,zhu} with a consequent lowering of the 
critical temperature.

The basic experimental result is the modulation of the (normalized) critical
current as a function of the applied gate voltage, as shown in
Fig.\ 2.
The input needed for the theory,
is the critical temperature $T_c$ as a function of the distance
from the boundary plane, and as a function of the applied field.
The theory then gives a prediction for the critical
current. 
Detailed work on the effect of doping on the critical
temperature of the cuprates \cite{Tallon,Jones} shows that 
T$_c$ follows
a parabolic relation: $ T/T_{c,max} = 1 - 82.6 * (n-0.16)^{2} $,
where n is the carrier concentration (holes per $CuO_2$ unit) 
and $T_{c,max}= 70K$ 
for the                             
material under study (Ca$_{0.3}$Sm$_{0.7}$Ba$_2$Cu$_{3}$O$_{y}$) \cite{Dong1}. 
From the specific characteristics of the specimen, given that     
a grain boundary of $24^0$ corresponds to a weakly-coupled bicrystal,
then the dependence of the concentration n on the distance follows
an exponential function in accordance with the previous work
\cite{BetourasJoynt}. So at zero applied gate voltage :
$ n(x) = 0.210- 0.206 *\exp (-0.2 x/ \xi) $,
where  $ \xi $ is the superconducting coherence length
which has an approximate value of $2nm$.

Also, the effect of the applied gate voltage $V_{g}$ is taken as
a linear contribution in the concentration function, since the 
induced charge density $\Delta$N for the specific material can
be found from the observation that  ${\Delta}N/V_{g} = C_{g}/|e| \simeq 3.2*
10^{11} cm^{-2}V^{-1}$ where $C_g$ is the areal gate capacitance
and $e$ is the electron charge. The contribution  is then 
$\delta n= 7*10^{-5} Volts^{-1} V_{g}$.

In the one-dimensional case we consider, the GL free energy 
that has to be minimized takes the form (we use a gauge
where the vector potential $\vec{A}=0$ for convenience,
since no magnetic field is present and
the free energy is gauge invariant):

\begin{eqnarray}
F= \alpha(x,T) {\Psi}^2+ \frac{\beta}{2} {\Psi}^4 - \frac{\hbar^2}{2m^{*}} 
(\frac{d}{dx}\Psi)^2,
\label{eq:free}
\end{eqnarray}
where $m^*$ is the effective mass.
We take the following form for $\alpha(x,T)$:

\begin{eqnarray}
\alpha(x,T) &=& \alpha_{0}{T_c(x)}^2 [\tanh(3.0 \sqrt{ T_c(x)/T - 1}) ]^2 
\;\;\; if\; T_c > T \\
            &=& 9 \alpha_{0}{T}^2 (T_c(x)/T - 1) 
\;\;\;\;\;\;\;\; if\;T_c < T.
\end{eqnarray}

This expression for $\alpha(x,T)$ for $T<T_c$ is an analytic
fit to the strong-coupling form of the gap function
in Bardeen-Cooper-Schrieffer theory.  Above $T_c$, we have less
knowledge about the form of the coefficients of
of GL theory.  We have chosen a form for
$\alpha$ in this regime guided by two considerations:  the
expression for $\alpha$ should continue smoothly
through $T_c$, and should be consistent with the 
equation $\xi= (\hbar^2/ 2 m^* \alpha)^{1/2}$,
where $\xi$ is the coherence length.
A function different from the usual first order
term  ($T-T_c$) in the expansion close to $T_c$ is necessary
in order to cover the whole  range of temperatures (it is stressed
here that $T_c$ varies with the distance). It is obvious that
if we expand the above function close to $T_c$ we get the usual
$(T-T_c)$ term.  $\beta$ is taken as
constant.
The theory thus involves the assumption that 
GL theory may be used over a broad range of temperatures. 
The great virtue of using this approach instead of a microscopic
theory is that the parameters can
be related to a number of observable quantities.
The above choice has been tested in the succesful fitting or 
prediction of several  quantities (calculation of the order parameter, 
NMR studies, specific heat etc.)\cite{QPLI,EurLett}.   

It is now convenient to write: $ \Psi= |\Psi| \exp(i\phi) $
in which case the current density $J$ is given by:
$ J= (\hbar e/m^*)* |\Psi|^2 * d\phi/dx $

These expressions may be simplified by the definitions:
 $ f(x) = \Psi/\Psi(\infty) $ ,
$ h(x) = \alpha(x,T)/\alpha(\infty,T) $
and $ j = J/J_{c}(\infty) $,
while x is taken in units of $ \xi $.

Here $J_c(\infty)$ is the bulk depairing current :
 \begin{equation}
J_{c}(\infty)= 2e |\Psi(\infty)|^2\frac{2}{3}(\frac{2\alpha(\infty,T)}
{3m^{2}})^{1/2}.
\end{equation}

The Euler-Lagrange equation corresponding to the GL free energy
in Eq.\ \ref{eq:free} is the differential equation:

\begin{equation}
      \frac{d^{2}f}{dx^{2}} - (4j^2/27f^3) + h(x) f- f^3 = 0,
\label{eq:diff}
\end{equation}

The computational problem 
is to solve the nonlinear differential equation (13)
with the boundary conditions: $ f(\pm \infty)  = \sqrt y $ and
  $ f'(\pm \infty)  = 0 $ ,
where y is the solution of the equation :
     $  y^2 - y^3 =(4/27) j^2 $.

At low current densities $j$, a superconducting solution exists.
The critical current is found by 
increasing $j$ until no superconducting solution exists.
This yields the quantity 
$\Delta{I_c}/I_{c0}$ where $\Delta{I_c}=I_c(V_g)-I_c(0)=I_c(V_g)-I_{c0}$.
We plot the
calculated results together with the experimental data in Fig.\ 2.
It's clear that the model under study can  reproduce the basic
experimental finding of the large electric field effect on grain
boundary Josephson junctions. 
The results show modulation in critical
current from a few percent for small voltages to almost 20\% for the largest
values that have been used. 

The calculation of the order parameter gives a very good picture 
of the suppression of superconductivity as we approach the boundary (Fig.\ 3).
Interestingly, there is also a region with order parameter values $|\Psi|$
greater than the bulk value of $\Psi$ due to the shape of the
function $T_c(x)$, leading to an enhancement of the superconductivity just
before the surpression. Temperature plays a more important role
in the shape of the order parameter, so at higher experimental temperature
there is a well defined nonsuperconducting region close to 
the grain boundary. The application of the gate voltage alters
the critical current which has small effect on the order parameter
(due to the fact that it is several order of magnitude less than
the depairing current at infinity). This small effect can be barely
detected in the asymptotic values of $\Psi$ as well as the value of
the peak of $\Psi$.
Furthermore, similar calculations in the underdoped regime 
(e.g. in an as-made $YBa_2Cu_3O_y$ film on bicrystal) have been 
performed and the quantity $\Delta{I_c}/I_{c0}$ shows that this case
is less sensitive to the field (about half the modification
observed in the overdoped regime).  

The calculations demonstrate that if there is 
relatively weak coupling between the two sides of the
boundary, a relatively small modification of the carrier density 
can have dramatic consequences.  The boundary serves effectively 
as a proximity-effect junction which changes from $S-S'-S$ towards
$S-N-S$ as the field is applied.  The result is the observed large field
effect. Thin films do not show the same effect
because the field is being applied to a strong superconducting region.

Another point to emphasize is that the above model
doesn't distinguish between s-wave or d-wave symmetry of the gap
function and consequently of the order parameter. The GL
theory has the identical form for the two cases \cite{QPLI}.
Thus the actual microscopic mechanism of high-$T_c$ 
superconductivity doesn't affect our results. 
 
These results provide guidance for further investigations in 
this field. The large field effect in JoFETs
can be accurately predicted within the Ginzburg-Landau theory 
while the complications of the microscopic theory are avoided.

This work was supported by the NSF  Grant. No. DMR-9214707
(J.B. and R.J.), by NSF Grant. No. DMR-9404579 (Z.-W.D. and
T. V.) and by the Dutch Program for High Temperature Superconductivity
(P.H.).


\begin{figure}
\caption{Schematic view of JoFET and the circuit outline for the field
effect measurements}
\label{one}
\end{figure}

\begin{figure}
\caption{ In Fig.2(a) the experimental data from Ref.[2]  
(the circles represent the
15 $\mu$m wide device and the squares the 30 $\mu$m wide device),
 in Fig.2(b) the results of the calculations from (GL) theory 
for the two experimental temperatures T=20K (solid line) and  T=4.2 K 
(dashed line) }
\label{two}
\end{figure}

\begin{figure}
\caption{ The calculated order parameter for the gate voltage of 40 Volts and
for the two experimental temperatures 4.2 K and 20 K }
\label{three}
\end{figure}

\end{document}